# Transverse photon spin beyond interfaces

Liang Peng[1+], Lingfu Duan[1], Kewen Wang[1], Fei Gao[2], Li Zhang[2], Gaofeng Wang[1], Yihao Yang[3], Hongsheng Chen[2#], Shuang Zhang[4*]

[1]*Key Laboratory for RF Circuits and Systems, Hangzhou Dianzi University,*

*Ministry of Education, Hangzhou, 310018, China*

[2] *State Key Laboratory of Modern Optical Instrumentation,*

*Zhejiang University, Hangzhou, 310027, China*

[3]*Division of Physics and Applied Physics, School of Physical and Mathematical Sciences,*

*Nanyang Technological University, Singapore 637371, Singapore*

[4]*School of Physics and Astronomy,*

*University of Birmingham*

E-mail: [+]pengl@hdu.edu.cn;

[#]hansomchen@zju.edu.cn;

[*]s.zhang@bham.ac.uk.

**Photons possess spin degree of freedom, corresponding to clockwise and counter clockwise rotating direction of the fields. Photon spin plays an important role in various applications such as optical communications, information processing and sensing. In conventional isotropic media, photon spin is aligned with the propagation direction of light, obeying spin momentum locking. Interestingly, at certain interfaces, the surface waves decaying away from the interface possess a photon spin transverse to its propagation, opening exciting opportunities for observation of spin dependent unidirectional excitation in confined systems. Here we propose and realize transverse photon spin (T-spin) in the interior of a bulk medium, without relying on the presence of any interfaces. We show the complete mapping of the T-spin of surface modes to that of the bulk modes by introducing the coupling between electric and magnetic responses**

**along orthogonal directions, i.e., the bianisotropy, into the medium. We further discover that an interface formed by two bianisotropic media of opposite orientations supports edge-dependent propagating modes with tunable cutoff frequencies. Our results provide a new platform for manipulating the spin orbit interaction of electromagnetic waves.**

## Introduction

In recent years, spin Hall effect (SHE) of light has attracted growing attention, [1-11] because of its great potential in facilitating novel nanophotonic and quantum applications. [12-18] As counterparts of SHE in solid-state physics, spin-orbital interaction of light produces spin-dependent transport of photons in systems under time reversal symmetry (T-symmetry). [1,2] In free space, the spin angular momentum (SAM) of the left-handed circular polarized (LCP) or right-handed circular polarized (RCP) is oriented along the direction of $\pm \bar{k}$, as constrained by the transversality condition $\bar{k} \cdot \bar{D}=0$ and $\bar{k} \cdot \bar{B}=0$. [19] Remarkably, the SAM of light in certain confined systems may be orthogonal to the propagation direction, as the so-called transverse spins (T-spins), whose E/H contains non-zero component longitudinal to $\bar{k}$. [20-22] Respecting the T-symmetry, the SAM of a T-spin changes its sign when propagation direction is reversed. This feature has been exploited in many important applications in the radio frequency and optical regimes, such as asymmetric scattering, directional spontaneous emission, spin-controlled radiation, and wave guiding. [23-27]

Constructing the T-spins requires the electric/magnetic field to rotate in the plane containing wave vector ($\bar{k}$), which does not usually occur for plane waves propagating inside conventional medium. Thus, T-spin was only realized for evanescent electromagnetic (EM) fields in some confined systems, with surface plasmonic polariton (SPP) as one of the most well-known examples. [2] For an SPP wave, its wavevector on each side of the interface can be expressed as $\bar{k}_{//} + i\bar{k}_T$ (with $\bar{k}_{//} \cdot \bar{k}_T = 0$), which, from the transverse wave condition, leads to a rotating field in the plane of wave propagation. The direction of the SAM is therefore perpendicular to the propagation direction of the surface wave. However, the SPP based T-spins are very limited in broad application scenarios, wherein external confinements should be lifted.

For instance, chiral quantum optics requires for three-dimensional T-spin dynamics in unbounded systems.[1] However, presence of a T-spin in a medium without any confinements has not been discovered till now.

Here we show that T-spins can exist in bulk state of the wave without involving evanescent component. Specifically, we propose a physical mapping of T-spin from surface modes to the bulk modes in homogeneous metamaterials. We show that the essential magneto-electric (ME) coupling along orthogonal directions (i.e. the bianisotropy) plays the same role as the evanescent wave vectors in generating the T-spin. Experimentally, we observe the bulk wave's T-spin and the corresponding asymmetric scattering. Further, we investigate the interesting wave guiding effect through T-spin cancellation on interfaces formed by two metamaterials with opposite T-spin bulk modes. The transport of such interface modes is shown to be highly robust against sharp bendings.

**T-spins beyond an Interface**

We start by reviewing the formation of T-spin in SPP waves. SPP wave is supported at the interface between a dispersive metal and air, as shown in Fig. 1a. It propagates along the interface with a complex wave vector, i.e. $\bar{k}_{spp} = k\hat{x} + i\sigma\hat{y}$ where $k$ and $\sigma$ are real. Accordingly, the E and H field of the SPP can be written as $\bar{E} = \hat{x}E_x + \hat{y}E_y$ and $\bar{H} = \hat{z}H_z$. On the air side, the divergence of $\bar{D}$ vanishes, leading to $kE_x + i\sigma E_y = 0$. Thus, there is a phase difference of $\pm\pi/2$ (depends on the sign of $\sigma$) between $E_x$ and $E_y$, resulting in a rotating E field in the $x$-$y$ plane, i.e. a T-spin along $z$ direction.

Suppose such a T-spin can also be excited inside a homogenous medium without the presence of an interface, as shown in Fig. 1b. Without losing generality, assuming the bulk mode is propagating along $x$ direction, i. e. $\bar{k}_{bulk} = k\hat{x}$. A direct mapping from the transverality condition $\bar{k} \cdot \bar{D}=0$ in the bulk medium to that of SPP $k(E_x + i\frac{\sigma}{k}E_y) = 0$ requires that

$D_x = \varepsilon_0(E_x + i\frac{\sigma}{k}E_y)$ with $\varepsilon_0$ being the free space constant. The above equation indicates that the medium shall be anisotropic with $\varepsilon_{xy} = i\frac{\sigma}{k}\varepsilon_{xx}$ in the absence of bianisotropy, i.e. there is no coupling between electric and magnetic responses. However, this constitutive relation does not exist in those lossless and reciprocal media, since $\bar{\bar{\varepsilon}}$ should be real and symmetric.

Interestingly, this problem can be solved if the medium possesses bianisotropy. Applying $\nabla \times \bar{E} = i\omega\bar{B}$ to the SPP wave, one obtains $kE_y = i\sigma E_x + \omega\mu_0 H_z$. Hence $\bar{k}_{spp} \cdot \bar{D} = 0$ can be rewritten as $k[(1-\frac{\sigma^2}{k^2})E_x + \frac{i\sigma\omega\mu_0}{k^2}H_z] = 0$, which, mapped to the bulk medium, is equivalent to the constituent equation $D_x = \varepsilon_0[(1-\frac{\sigma^2}{k^2})E_x + \frac{i\sigma\omega\mu_0}{k^2}H_z]$. The relationship between $H_z$ and $D_x$ indicates that the electric-magnetic coupling, i.e. bianisotropy, is essential for a bulk medium to support T-spins. Thus, a complete mapping of the T-spin from an SPP mode to a bulk mode is accomplished.

For a lossless and reciprocal bianisotropic medium, the constitutive relation can be written as $\bar{B} = \bar{\bar{\mu}} \cdot \bar{H} + i\bar{\bar{\chi}} \cdot \bar{E}$ and $\bar{D} = \bar{\bar{\varepsilon}} \cdot \bar{E} - i\bar{\bar{\chi}}^T \cdot \bar{H}$. Without losing generality, we assume $\bar{\bar{\varepsilon}}$ and $\bar{\bar{\mu}}$ take the form $\bar{\bar{\varepsilon}} = diag[\varepsilon_x, \varepsilon_y, \varepsilon_z]$ and $\bar{\bar{\mu}} = diag[\mu_x, \mu_y, \mu_z]$, and the only nonzero element in $\bar{\bar{\chi}}$ is $\chi_{zx}$. For the TM polarization, the electric field of the wave and the dispersion satisfy $\bar{E} = [\hat{x}\frac{-k_y}{\omega\varepsilon_x} + \hat{x}\frac{i\chi_{zx}}{\varepsilon_x} + \hat{y}\frac{k_x}{\omega\varepsilon_y}]e^{ik_x x + ik_y y}$ and $\frac{\varepsilon_x}{\varepsilon_y}k_x^2 + k_y^2 = \omega^2(\varepsilon_x\mu_z - \chi_{zx}^2)$, respectively. In particular, with $k_y = 0$, the electric field acquires an additional component proportional to $\chi_{zx}$ along $x$ direction, which is $\pi/2$ out of phase relative to its $y$ component. Thus, the $x$ and $y$ components form a rotating electric field in the $x$-$y$ plane as shown in Fig. 1c. Mathematically, the transverse SAM is expressed as $\bar{S} = -\hat{z}\varepsilon_0\chi_{zx}k_x/(2\omega^2\varepsilon_x\varepsilon_y)$, which further confirms that ME coupling ($\chi_{zx}$) plays the key role in forming the T-spin of bulk modes, in a way very similar to the imaginary wavevector for the SPPs. In Fig. 1d, we show the spatial dispersion of the T-spin for the eigenmodes of a bianisotropic metamaterial, in which a continuous transition from $\hat{S}\cdot\hat{z}>0$ to $\hat{S}\cdot\hat{z}<0$ is observed.

**T-spin Dynamics in Realistic Metamaterial**

In the experiments, we adopt an unbalanced split-ring resonator (USRR) structure as the building block to construct the bulk metamaterial. The detailed information of the metamaterial and the USRR structure are shown in Fig. 2a and Fig. 2b. The unit cell lattice constants in $x$, $y$ and $z$ directions are 12mm, 12mm, and 5mm, respectively. Two identical USRRs made of copper are fabricated on both sides of 1mm thick FR4 substrate (a printed-circuit board, whose dielectric constant is 4.3 with loss tangent 0.025). For more details of the design, please refer to the supplementary information section III.

For the TM-type mode that we are interested in, only four effective parameters are involved, i.e., $\varepsilon_x$, $\varepsilon_y$, $\mu_z$ and $\chi_{zx}$, whose effective values extracted from the transmittance and reflectance by a slab made of one-layer USRRs, through the parameter-retrieval approach (Fig. 2c).[28, 29] The forbidden band can be determined from the retrieved parameters, as indicated as the shadow region in Fig. 2c and Fig. 2d. In Fig. 2d, we present the refractive index of the mode with and without the bianisotropy term. It is seen that the medium would be transparent without $\chi_{zx}$ in the forbidden band of the metamaterial. Thus, it is clear that the forbidden band arises from the contribution of the ME coupling.

In order to measure the T-spin of the propagating mode, we carry out near field measurement across the sample surface. In the measurement, a bulk metamaterial composed of 24×12×10 USRR unit cells is fabricated. The TM mode is excited inside the bulk through a feeding waveguide at the center of bulk surface perpendicular to $x$ direction. The E-field distribution is measured at a distance 5mm above the top surface of the bulk medium (For the detailed configurations, please refer to the supplementary information section IV). The measured distributions of $E_x$ and $E_y$ of the TM mode at 3.6 GHz are shown in Fig. 2e. It is clearly seen there exists $-\pi/2$ phase difference between $E_x$ and $E_y$, i.e., the electric field rotates in the plane containing $\bar{k}$, and hence the presence of transverse SAM is experimentally confirmed. Fig. 2f shows the simulation results on the polarization state. The simulation shows that the field is strongly inhomogeneous due to the localized resonance in individual USRRs, while this local inhomogeneity is not present in the measurement. This is because the probe has a finite size, which records locally averaged field. Nonetheless, the overall phase patterns for both electric field components are in good agreement between the simulation and the experiment.

The presence of T-spin in the bulk mode supported by the bianisotropic metamaterial provides opportunity for spin-dependent asymmetric excitation of the mode. In Fig. 3a, we show the schematic of the excitation of the bulk mode, where a beam of circular polarization is incident onto the sample edge to excite the bulk wave propagating along $+\hat{x}$ direction. From previous analysis, the T-spin excited into the bulk ($\hat{k} = +\hat{x}$) would have its SAM pointing to the $+\hat{z}$ direction, which matches that of LCP incidence. In Fig. 3b, we show the power collected by the receiver located at the right-hand side of the bulk. Both the simulated and measured results show that the detected signal is strong with LCP incidence, with the peak power observed around 4GHz, where the polarization matching between the LCP and the T-spin is optimized (for more details please refer to the supplementary materials). However, for the RCP incidence, the excitation is very weak due to the mismatch of the incident spin and the T-spin of the bulk mode. Thus, we directly confirm the spin-selective excitation of the bulk mode through the measurement.

**The Interface Modes**

The preceding theoretical analysis shows that a very large bianisotropy can result in insulating phase for photons. Interestingly, when two metamaterials with opposite bianisotropy form an interface, there exist interface state that shows robust propagation, though this robustness is not of topological origin.

Assume an interface is formed by two metamaterials with opposite bianisotropy while all other parameters being the same ($\bar{\bar{\varepsilon}}_1 = \bar{\bar{\varepsilon}}_2$, $\bar{\bar{\mu}}_1 = \bar{\bar{\mu}}_2$ and $\bar{\bar{\chi}}_1 = -\bar{\bar{\chi}}_2$), as shown in Fig. 4a. Due to the fact that the metamaterials do not exhibit $z$ rotational symmetry in the $x$-$y$ plane, the interface mode depends on the orientation of the interface. Without losing generality, we assume the interface is in the $x$-$z$ plane while the orientation of the bulk metamaterial (represented by the local coordinate $x'$-$y'$) forms an angle $\phi$ with the interface, as shown in the insets of Fig. 4a. By solving the boundary problem (Supplemental Materials), we obtain

$$k_y^\pm = \frac{-k_x(\varepsilon_x - \varepsilon_y)\sin\phi\cos\phi \pm i\omega\chi\varepsilon_y\cos\phi}{\varepsilon_x\sin^2\phi + \varepsilon_y\cos^2\phi}, \quad (2)$$

with $k_x = \omega\sqrt{\mu_z(\varepsilon_x\sin^2\phi + \varepsilon_y\cos^2\phi) - \chi^2\sin^2\phi}$. Eq. (2) shows that the propagation of the interface modes can be manipulated through tuning the two bulk media's orientation angle ($\phi$). In particular, for $\varepsilon_x\mu_z < \chi^2$, with $\phi=0$ and $\phi=\pm\pi$, $k_x$ is real; but with $\phi=\pm\pi/2$, both $k_x$ and $k_y$ become imaginary (or zero). Because the imaginary part of $k_y$ has to be positive to guarantee the decay of interface modes in the $y>0$ region, the interface modes can only exist within a limited range of $\phi$. For given constitutive parameters, the dispersion and in particular the cut-off of the surface mode are $\phi$-dependent (supplementary material). In Fig. 4b, we show the dispersion curves of these interface states with different $\phi$. A variation of the cutoff frequency for the interface modes by tuning $\phi$ is clearly observed. For $\phi=0$, the $H_z$ distribution at 3GHz is shown in Fig. 4c, wherein well-confined interface mode is observed. However, with $\phi=\pi$, according to Eq. (2), no surface modes could exist. In the experiments, the transmittance of the interface modes with $\phi=0$ and $\phi=\pi$ is measured and shown in Fig. 4d. It is observed that the transmission with $\phi=0$ occurs above 3GHz, which is consistent with our numerical evaluations in Fig. 4b. However, with $\phi=\pi$ no efficient transmission is observed, which again confirms our theoretical analysis.

The measured transmittances for the interface mode with different number of bends are shown in Fig. 4f. It is observed that the transmission is very efficient for frequency above 3.2GHz, which agrees with our numerical simulation in Fig. 4b. Importantly, the transmittances with different number of bends are almost the same, indicating highly robust transmission of the edge mode. And, the robustness of the zero-indexed transmission is verified by the transmittances shown in Fig. 4f. It is interesting to note that in contrast to possession of T-spin in the bulk mode, it is found that the interface mode is linearly polarized, as shown in Fig. 4a. The linear polarization of the surface waves is enforced by the boundary condition between two media of opposite bianisotropy (Supplementary Materials). This also indicates the exact cancellation of the transverse spin from the contribution of the evanescent wave vector and that of bianisotropy.

**Conclusions**

In conclusion, we have demonstrated T-spin for bulk modes inside a medium with orthogonal ME coupling. A direct mapping from SPP to bulk modes shows that the ME coupling in a bulk medium can act equivalently to the evanescent wave vector of a surface wave in forming the T-spin. The T-spin is directly confirmed by both the experimental measurement and numerical simulations. We further show the spin selective excitation of the bulk mode, T-spin induced insulating phase and the corresponding edge mode. In contrast to previously studied T-spin in evanescent waves, our work has provided a new approach to construct the T-spins in the bulk mode of metamaterials. Our work may facilitate many photonic applications, such as spin-selective absorption and extinction, polarization detection, asymmetric emission and guidance, as well as on-chip manipulation of photons.

**Acknowledgements**

This work is financially supported by the Natural Science Foundation of China (NSFC) under Grant 61875051 and 61372022. S. Z. acknowledges the support from ERC Consolidator Grant (TOPOLOGICAL).

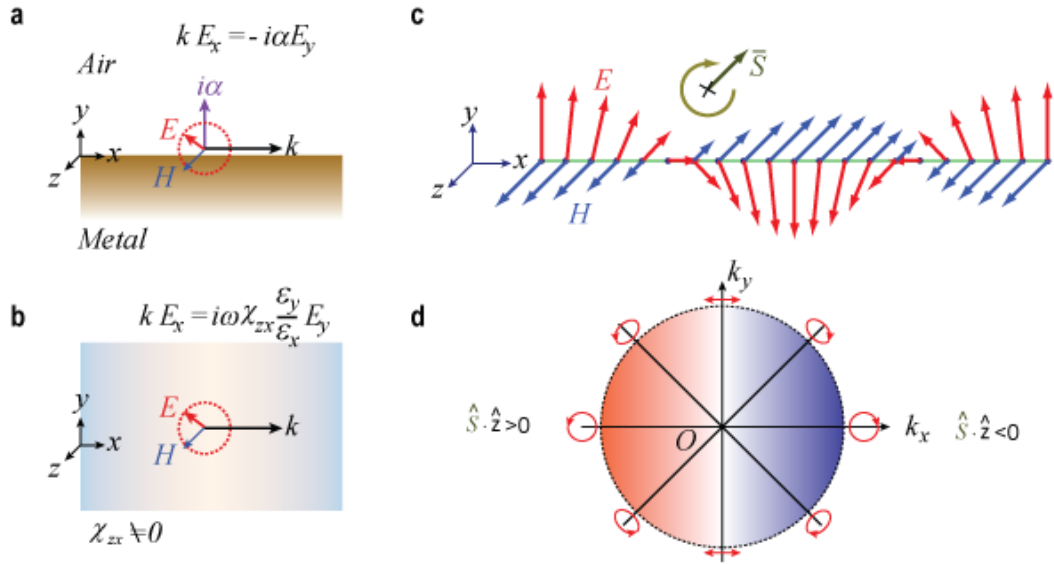

**Figure 1. T-spins in the bulk mode of a homogeneous material. a**, T-spin is excited due to the imaginary wavevector for the SPPs. **b**, T-spin is excited due to the non-zero ME coupling ($\chi_{zx} \neq 0$). **c**, T-spin dynamics of light ($\chi_{zx} > 0$). In the calculation, $k_y = 0$ is assumed. **d**, The spatial dispersion of the T-spin in bulk metamaterials.

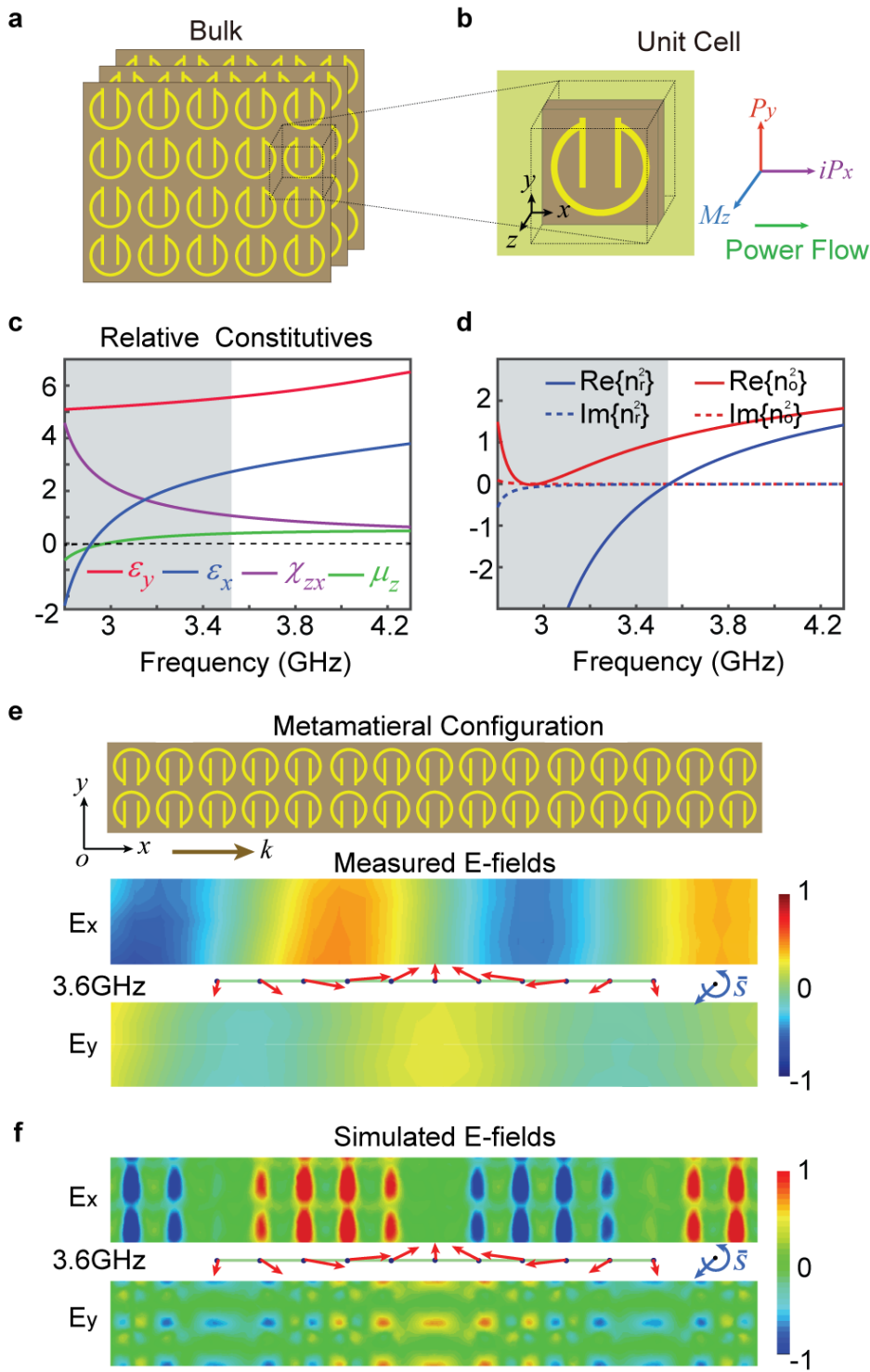

**Figure 2. Demonstraton of T-spin in a metamaterial with orthogonal ME feature. a**, The metamaterial configuration, which consists of three dimensional array of SRRs. **b**, Polarization and magnetization on an individual USRR for a wave propagating along $x$ direction. **c**, The retrieved effective parameters ($\varepsilon_x$, $\varepsilon_y$, $\mu_z$ and $\chi_{zx}$) relevant to the TM-type mode. The shadow region represents the forbidden band for the TM-type modes. The imaginary parts of these parameters are negligible and not shown. **d**, Numerically calculated dispersion curves for $n_o^2 = \varepsilon_x \mu_z$ and $n_r^2 = \varepsilon_x \mu_z - \chi_{zx}^2$. It is seen that from below 3GHz to above 3.5GHz, $n_o^2$ remains real and positive, but $n_r^2$ is negative, which explains the T-spin induced insulating effect. **e**, The experimentally measured electric field distribution at 3.6GHz. **f**, The simulated electric field distribution in the metamaterial at 3.6GHz. In **e** and **f**, the rotation of the electric field and the SAM are schematically shown for better interpretation.

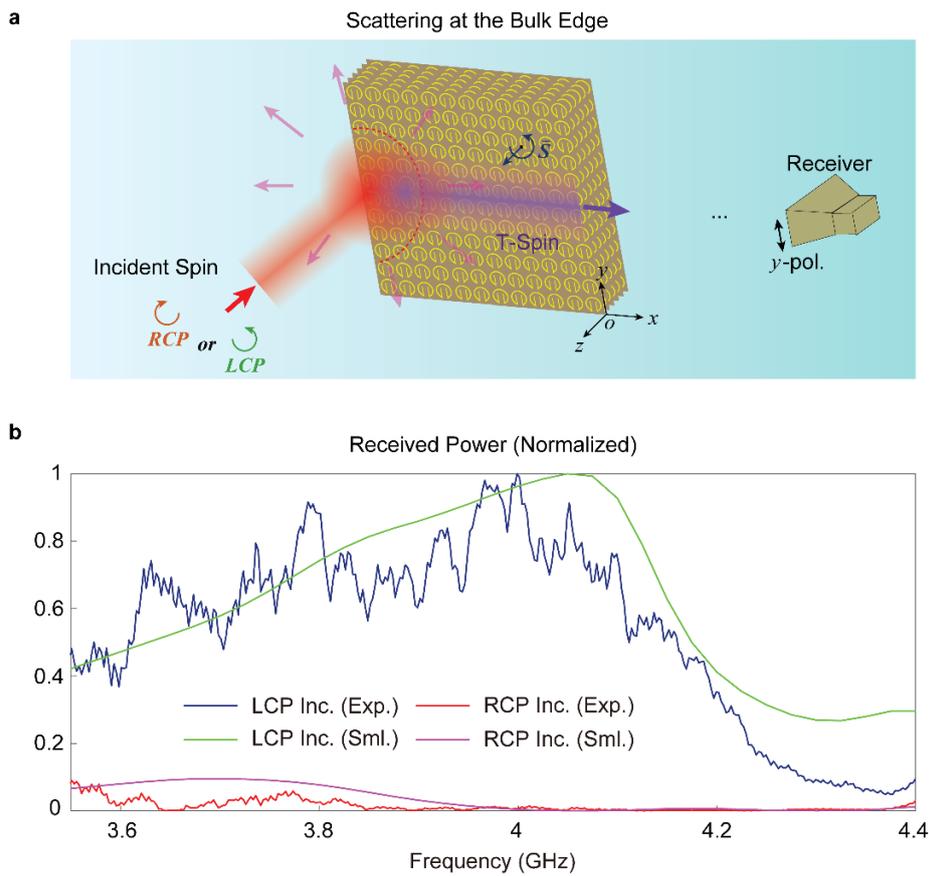

**Figure 3. Polarization locked scattering from finite-sized USRR arrays. a**, Schematic of scattering from a finite-sized bulk metamaterial made of $12 \times 12 \times 6$ USRRs. In the experiments, the power receiver is placed on the right-hand side. **b**, The received power spectrum in both the measurements and simulations. The results are both normalized.

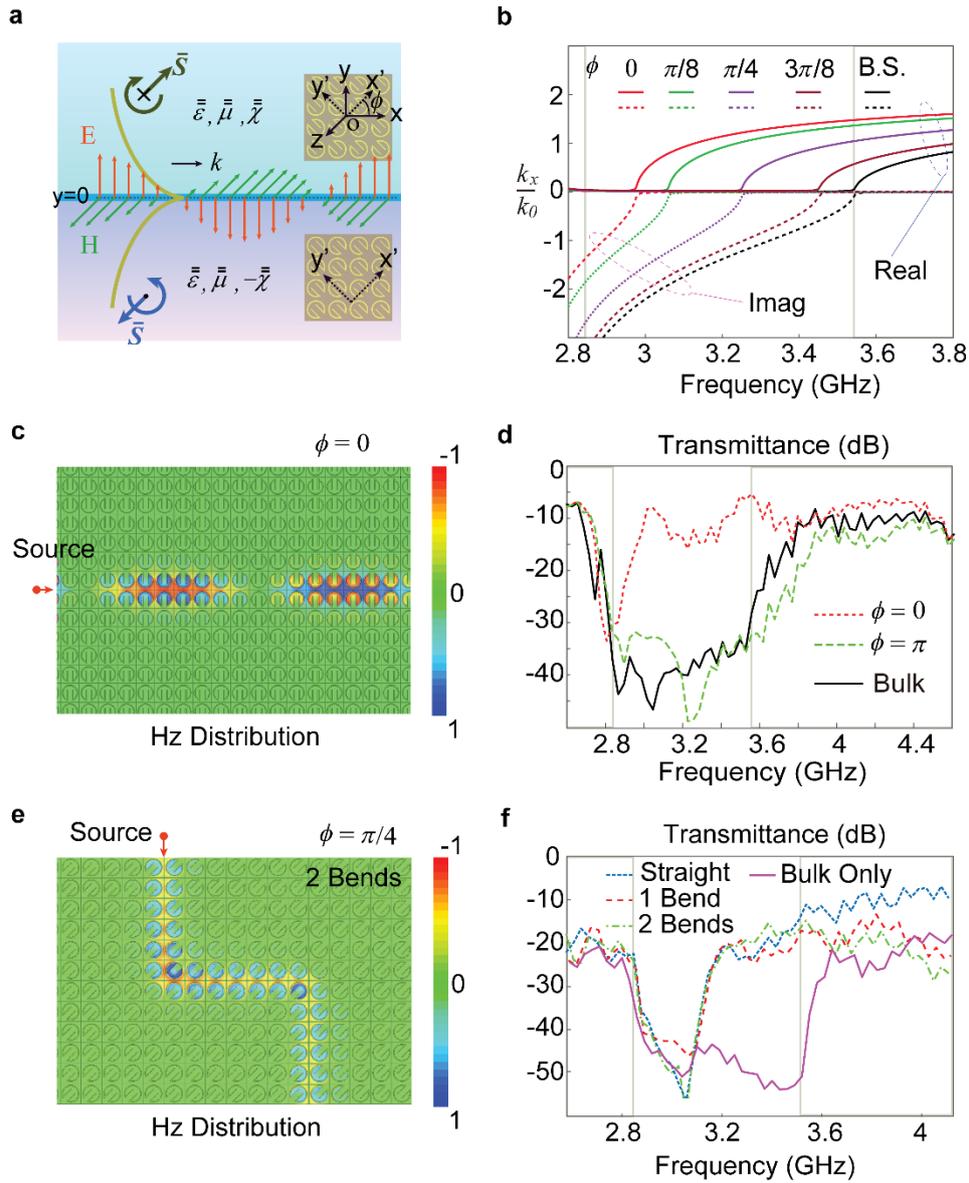

**Figure 4. Compensation of T-spins on an edge and propagation of interface modes. a**, Linearly polarized modes on an interface between two bianisotropic bulks with opposite orientations. **b**, Numerically calculated dispersion for the interface mode. The upper and lower bounds for the forbidden band are indicated. **c**, Simulated H-field distribution for the interface mode with $\phi = 0$. **d**, Measured transmittance for $\phi = 0$ and $\phi = \pi$. For better comparison, the transmittance for the bulk metamaterial is also shown. **e**, Simulated H-field distribution for the $\phi = \pi/4$ interface mode with two-bends passage at 3.2GHz. Because $k_x$ is a small, there is almost no back-reflection. **f**, Measured transmittance for the $\phi = \pi/4$ interface mode for interface configurations with different number of bends but the same length. The detailed interface configurations can be found in the supplementary information.